# An Improved UP-Growth High Utility Itemset Mining

Adinarayanareddy B
M.Tech (CSE)
Dept. of CSE
JNTUK – UCEV
Vizianagaram, India

O Srinivasa Rao, PhD.
Assistant Prof. of CSE & Head
Dept. of CSE
JNTUK – UCEV
Vizianagaram, India

MHM Krishna Prasad, PhD.
Associate Prof. of CSE & Head
Dept. of I T
JNTUK – UCEV
Vizianagaram, India

## ABSTRACT
Efficient discovery of frequent itemsets in large datasets is a crucial task of data mining. In recent years, several approaches have been proposed for generating high utility patterns, they arise the problems of producing a large number of candidate itemsets for high utility itemsets and probably degrades mining performance in terms of speed and space. Recently proposed compact tree structure, viz., UP-Tree, maintains the information of transactions and itemsets, facilitate the mining performance and avoid scanning original database repeatedly. In this paper, UP-Tree (Utility Pattern Tree) is adopted, which scans database only twice to obtain candidate items and manage them in an efficient data structured way. Applying UP-Tree to the UP-Growth takes more execution time for Phase II. Hence this paper presents modified algorithm aiming to reduce the execution time by effectively identifying high utility itemsets.

## 1. INTRODUCTION
Association rules mining (ARM) [6] is one of the most widely used techniques in data mining and knowledge discovery and has tremendous applications like business, science and other domains. Make the decisions about marketing activities such as, e.g., promotional pricing or product placements.

A high utility itemset is defined as: A group of items in a transaction database is called itemset. This itemset in a transaction database consists of two aspects: First one is itemset in a single transaction is called internal utility and second one is itemset in different transaction database is called external utility. The transaction utility of an itemset is defined as the multiplication of external utility by the internal utility. By transaction utility, transaction weight utilizations (TWU) can be found. To call an itemset as high utility itemset only if its utility is not less than a user specified minimum support threshold utility value; otherwise itemset is treated as low utility itemset.

To generate these high utility itemsets mining recently in 2010, UP-Growth (Utility Pattern Growth) algorithm [11] was proposed by Vincent S. Tseng et al. for discovering high utility itemsets and a tree based data structure called UP-Tree (Utility Pattern tree) which efficiently maintains the information of transaction database related to the utility patterns. Four strategies (DGU, DGN, DLU, and DLN) used for efficient construction of UP-Tree [11] and the processing in UP-Growth [11]. By applying these strategies, can not only efficiently decrease the estimated utilities of the potential high utility itemsets (PHUI) but also effectively reduce the number of candidates. But this algorithm takes more execution time for phase II (identify local utility itemsets) and I/O cost.

In this paper, the existing UP-Growth algorithm is improved to generate high utility itemsets efficiently for large datasets and reduce execution time in phase II compared with existing UP-Growth algorithm. In the experimental section, experiments are conducted on our improved algorithm and existing algorithm with variety of synthetic and real-time datasets.

## 2. RELATED WORK
Association rule mining is considered to be an interesting research area and studied widely [1-9] by many researchers. In the recent years, some relevant methods have been proposed for mining high utility itemsets from transaction databases.

In 1994, Agrawal .R et al. [1] proposed Apriori algorithm by exploit "downward closure property", which is the pioneer for efficiently mining association rules from large databases. This algorithm generated and tested candidate itemsets iteratively. This may scan database multiple times, so the computational cost is high. In order to overcome the disadvantages of Apriori algorithm and efficiently finds frequent itemsets without generating candidate itemsets, a frequent pattern Growth (FP-Growth) is proposed by Han .J et al. [5].

The FP-Growth was used to compress a database into a tree structure which shows a better performance than Apriori. Although it has two limitations: (i). It treats all items with the same price. (ii). In one transaction each item appears in a binary (0/1) form, i.e. either present or absent. In the real world, each item in the supermarket has a different prices and single customer may take same item multiple times. Therefore, finding only traditional frequent patterns in a database cannot fulfill the requirement of finding the most valuable customers/itemsets that contribute the most to the total profit in a retail business. Later different algorithms proposed like Two-Phase [7], IIDS [6] and IHUP [2].

In 2006, H. Yao et al. proposed UMining [8] algorithm to find almost all the high utility itemsets from an original database. But it suffers to capture a complete set of high utility itemsets. Later, In 2010 V. S. Tseng et al. [11] proposed UP-Growth algorithm to rectify the problems of FP-Growth.

## 3. PROPOSED METHOD
The goal of utility mining is to discover all the high utility itemsets whose utility values are beyond a user specified threshold in a transaction database.

## 3.1 UP-Growth Algorithm

The UP-Growth [11] is one of the efficient algorithms to generate high utility itemsets depending on construction of a global UP-Tree. In phase I, the framework of UP-Tree follows three steps: (i). Construction of UP-Tree [11]. (ii). Generate PHUIs from UP-Tree. (iii). Identify high utility itemsets using PHUI.

The construction of global UP-Tree [11] is follows, (i). Discarding global unpromising items (i.e., DGU strategy) is to eliminate the low utility items and their utilities from the transaction utilities. (ii). Discarding global node utilities (i.e., DGN strategy) during global UP-Tree construction. By DGN strategy, node utilities which are nearer to UP-Tree root node are effectively reduced. The PHUI is similar to TWU, which compute all itemsets utility with the help of estimated utility. Finally, identify high utility itemsets (not less than min_sup) from PHUIs values. The global UP-Tree contains many sub paths. Each path is considered from bottom node of header table. This path is named as conditional pattern base (CPB).

## 3.2 Improved UP-Growth

Although DGU and DGN strategies are efficiently reduce the number of candidates in Phase 1(i.e., global UP-Tree). But they cannot be applied during the construction of the local UP-Tree (Phase-2). Instead use, DLU strategy (Discarding local unpromising items) to discarding utilities of low utility items from path utilities of the paths and DLN strategy (Discarding local node utilities) to discarding item utilities of descendant nodes during the local UP-Tree construction. Even though, still the algorithm facing some performance issues in phase-2. To overcome this, maximum transaction weight utilizations (MTWU) are computed from all the items and considering multiple of min_sup as a user specified threshold value as shown in algorithm. By this modification, performance will increase compare with existing UP-Tree construction also improves the performance of UP-growth algorithm. An improved utility pattern growth is abbreviated as IUPG.

## IUPG-Algorithm:

**Input:** Transaction database D, user specified threshold.
**Output:** high utility itemsets.

## Begin

1. Scan database of transactions $T_d \in D$
2. Determine transaction utility of $T_d$ in D and TWU of itemset (X)
3. Compute min_sup (MTWU * user specified threshold)
4. If (TWU(X) ≤ min_sup) then Remove Items from transaction database
5. Else insert into header table H and to keep the items in the descending order.
6. Repeat step 4 & 5 until end of the D.
7. Insert $T_d$ into global UP-Tree
8. Apply DGU and DGN strategies on global UP- tree
9. Re-construct the UP-Tree
10. **For** each item $a_i$ in H do
11. Generate a PHUI Y= X U $a_i$
12. Estimate utility of Y is set as $a_i$'s utility value in H
13. Put local promising items in Y-CPB into H
14. Apply strategy DLU to reduce path utilities of the paths
15. Apply strategy DLN and insert paths into $T_d$
16. If $T_d \neq$ null then call for loop

## End for
## End

## 4. EXPERIMENTAL EVALUATION

In this section, experimental results on synthetic datasets and real world databases [10] are summarized on both UP-Growth and Improved UP-Growth algorithm. These experiments were conducted on 2.53 Intel(R) Core(TM) i3 Processor with 2 GB of RAM, and running on Windows 7 operating system. All algorithms were implemented in java language (JDK1.5) and applied both synthetic and real datasets to evaluate the performance of the both algorithms.

## 4.1 Synthetic Dataset

First, the performance deviation of UP-Growth (UPG) is shown and Improved UP-Growth (IUPG) algorithms on the synthetic datasets T10I6D10K. Where *T* is the average size of transactions; I is the average size of maximal potential frequent itemsets; D is the total number of transactions and N is the number of distinct items. Table-1 shows the execution times on various min_sup values from 60% to 90%. Fig-1 and Fig-2 shows the performance evaluation of UPG and IUPG for phase I and Phase II execution times on various min_sup values from 60% to 90%.

**Table-1: Execution times on T10I6D10K**

| Dataset | UPG | EUPG | UPG | EUPG |
|---|---|---|---|---|
| Min_Sup (%) | phase I (sec) | | phase II (ms) | |
| 90 | 268 | 248 | 8 | 0 |
| 85 | 485 | 269 | 14 | 2 |
| 80 | 480 | 267 | 15 | 2 |
| 75 | 480 | 269 | 15 | 2 |
| 70 | 502 | 280 | 22 | 2 |
| 65 | 1003 | 280 | 35 | 8 |
| 60 | 1040 | 282 | 79 | 59 |

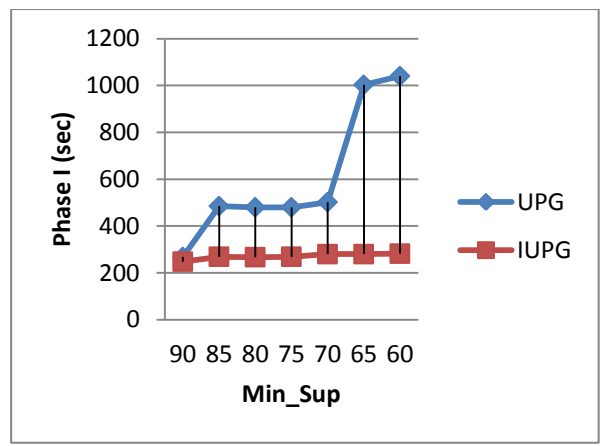

**Fig-1: T10I6D10K Phase-I Time**

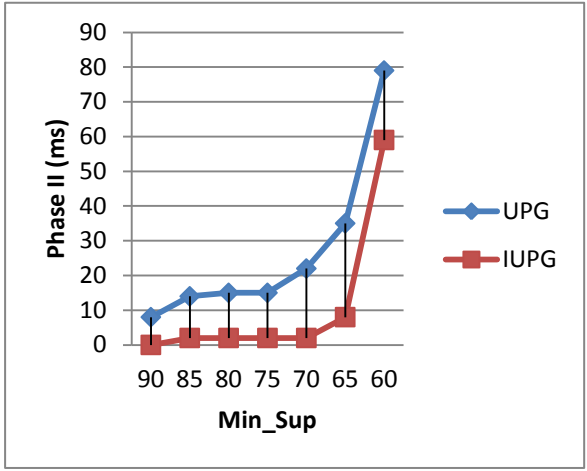

**Fig- 2: T10I6D10K Phase-II Time**

## 4.2 Real time Dataset

Here compare the performance of UPG and IUPG on real time chess dataset [10]. Table-2 shows the execution times on various min_sup values from 65% to 90%. Fig-3 and Fig-4 shows the performance evaluation of UPG and IUPG for phase I and Phase II execution times.

**Table-2: Execution times on Chess**

| Dataset | UPG | IUPG | UPG | IUPG |
|---|---|---|---|---|
| Min_Sup (%) | phase I (sec) | | phase II (ms) | |
| 90 | 17 | 14 | 19 | 19 |
| 85 | 28 | 15 | 28 | 18 |
| 80 | 31 | 16 | 37 | 24 |
| 75 | 34 | 19 | 43 | 28 |
| 70 | 33 | 21 | 48 | 30 |
| 65 | 36 | 28 | 57 | 33 |

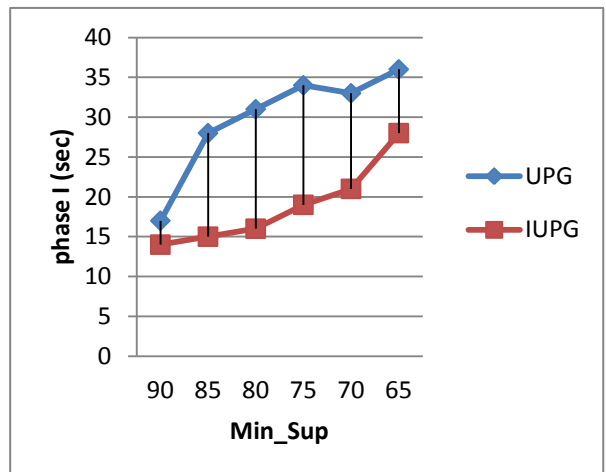

**Fig-3: Execution time for phase I on Chess**

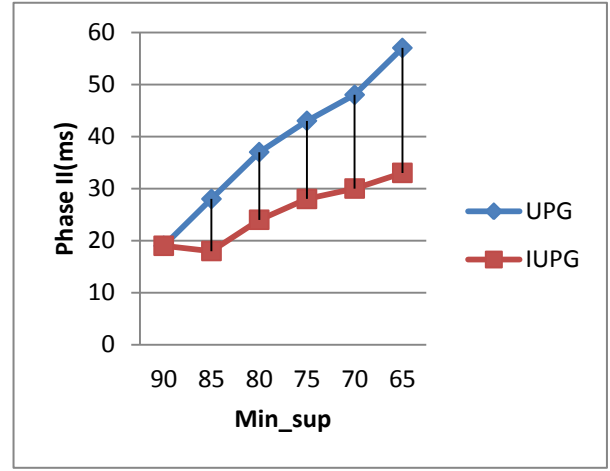

**Fig-4: Execution time for phase II on Chess**

## 4.3 Scalability

In this section, the size of T10I6 dataset is varied to evaluate the scalability for UPG and IUPG algorithms. In Table-3 shows the execution times on various dataset sizes and min_sup is 85%. However, the execution time of IUPG is less than UPG. When the database size increases, the execution time for identifying high utility itemsets also increases. Hence, UP-Growth algorithm requires more processing time than IUPG.

**Table-3: Execution times on Scalability**

| Dataset | UPG | IUPG | UPG | IUPG |
|---|---|---|---|---|
| Size | Phase- 1(sec) | | Phase - II(ms) | |
| 1000 | 26 | 25 | 8 | 2 |
| 5000 | 135 | 127 | 14 | 9 |
| 10000 | 328 | 269 | 26 | 16 |
| 25000 | 994 | 768 | 32 | 24 |
| 50000 | 2468 | 1958 | 67 | 36 |

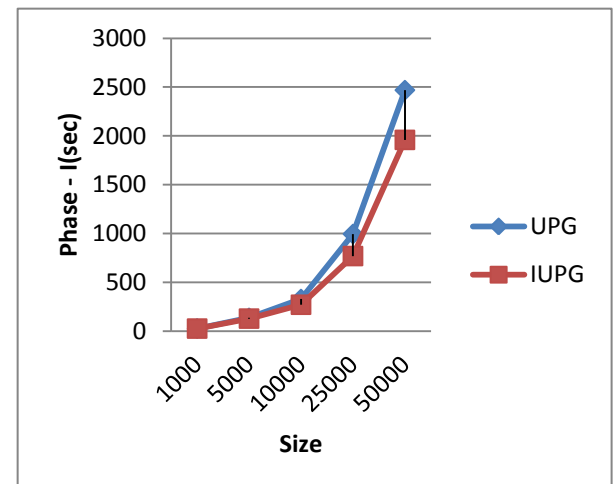

**Fig-5: Execution time for phase I on Scalability**

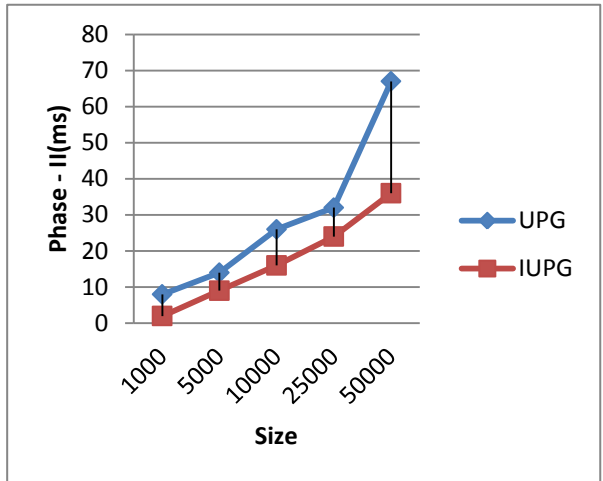

**Fig-6: Execution time for phase II on Scalability**

By the experimental results, the Improved UP-Growth is efficiently reducing the execution time of phase II and also effectively identifies the high utility itemsets on both synthetic and real datasets. Therefore, IUPG algorithm achieves better performance than UPG.

## 5. CONCLUSION AND FUTURE WORK

Mining high utility itemsets becomes more significant. In this paper, the Improved UP-Growth (IUPG) algorithm evaluated with Existing UP-Growth (UPG) algorithm. These algorithms are experimented on synthetic datasets and real time datasets for different support threshold. From the experimental observation, the conclusion is that, IUPG algorithm performs well than UPG algorithm for different support values. Also the IUPG algorithm scales well as the size of the transaction database increases. The future work would focus on the different issues to improve phase-I in terms of execution and memory space cost.

## 6. REFERENCES

[1] R. Agrawal and R. Srikant.: Fast algorithms for mining association rules. *In Proc. of the 20th Int'l Conf. on Very Large Data Bases, pp. 487-499, 1994.*

[2] C. F. Ahmed, S. K. Tanbeer, B. S. Jeong, and Y. K. Lee.: Efficient tree structures for high utility pattern mining in incremental databases. *In IEEE Transactions on Knowledge and Data Engineering, Vol. 21, Issue 12, pp. 1708-1721, 2009.*

[3] R. Chan, Q. Yang, and Y. Shen.: Mining high utility itemsets. *In Proc. of Third IEEE Int'l Conf. on Data Mining, pp. 19-26, 2003.*

[4] A. Erwin, R. P. Gopalan, and N. R. Achuthan.: Efficient mining of high utility itemsets from large datasets. *In Proc. of PAKDD 2008, LNAI 5012, pp. 554-561.*

[5] Jiawei. Han, Jian. Pei, and Y. Yin.: Mining frequent patterns without candidate generation. *In Proc. of the ACM-SIGMOD Int'l Conf. on Management of Data, pp. 1-12, 2000.*

[6] Y. C. Li, J. S. Yeh, and C. C. Chang.: Isolated items discarding strategy for discovering high utility itemsets. *In Data & Knowledge Engineering, Vol. 64, Issue 1, pp. 198-217, Jan., 2008.*

[7] Y. Liu, W. Liao, and A. Choudhary.: A fast high utility itemsets mining algorithm. *In Proc. of the Utility-Based Data Mining Workshop, 2005.*

[8] H. Yao, H. J. Hamilton, and L. Geng.: A unified framework for utility-based measures for mining itemsets. *In Proc. of ACM SIGKDD 2nd Workshop on Utility-Based Data Mining, pp. 28-37, USA, Aug., 2006.*

[9] S. J. Yen and Y. S. Lee.: Mining high utility quantitative association rules. In *Proc. of 9th Int'l Conf. on Data Warehousing and Knowledge Discovery, Lecture Notes in Computer Science 4654*, pp. 283-292, Sep., 2007.

[10] Frequent itemset mining implementations repository, http://fimi.cs.helsinki.fi/

[11] Vincent. S. Tseng, C. W. Wu, B. E. Shie, and P. S. Yu.: UP-Growth: An Efficient Algorithm for High Utility Itemset Mining. In Proc. of ACM-*KDD*, Washington, DC, USA, pp. 253-262, July 25–28, 2010.